\documentclass
[twocolumn,showpacs,preprintnumbers,amsmath,amssymb,prc]{revtex4}
\usepackage{graphicx}
\usepackage{dcolumn}
\usepackage{bm}

\begin{document}

\title{Conditions for equivalence of Statistical Ensembles in Nuclear Multifragmentation}

\author{ Swagata Mallik and Gargi Chaudhuri}

\affiliation{Theoretical Physics Division, Variable Energy Cyclotron Centre, 1/AF
Bidhan Nagar, Kolkata700064,India}

\begin{abstract}
Statistical models based on canonical and grand canonical ensembles
are extensively used to study intermediate energy heavy-ion collisions.
The underlying physical assumption behind canonical and grand canonical
models is fundamentally different, and in principle agree only in
the thermodynamical limit when the number of particles become infinite.
Nevertheless, we show that these models are equivalent in the sense
that they predict similar results if certain conditions are met even
for finite nuclei. In particular, the results converge when nuclear
multifragmentation leads to the formation of predominantly nucleons
and low mass clusters. The conditions under which the equivalence
holds are amenable to present day experiments.
\end{abstract}


\maketitle
\section{Introduction}
In the disintegration of a nuclear system formed
by the collision of two heavy-ions at intermediate energy, it is assumed
that a statistical equilibrium is reached. This facilitates the use
of statistical models \cite{Das1,Bondorf,Gross1} in order to obtain the
yields of the composites at the freeze-out volume. In such models of nuclear
disassembly the populations of the different channels are solely decided
by their statistical weights in the available phase space. One can use
different ensembles (microcanonical, canonical or grand canonical)
in order to account for the fragmentation of the nucleus into different
channels. The partitioning into available channels can be solved in
the canonical model \cite{Das1} where the number of particles in
the nuclear system is finite (as it would be in experiments). Even
when the number of particles is fixed one can replace a canonical
model by a grand canonical model where the particle number fluctuates
but the average number is constrained to a given value \cite{Chaudhuri1,Reif}.
Both canonical and grand canonical models have been extensively used
to study the physics of intermediate energy heavy ion collisions \cite{Das1,Bondorf,Randrup,Raduta}
and results for different observables have been routinely compared
to experimental data \cite{Botvina,Chaudhuri3,Chaudhuri4,Ogul,Barz}.

It is well known that results from different statistical ensembles agree
in the thermodynamical limit \cite{Reif}, that is, when the number
of particles become infinite. For example, for one kind of particle
(nucleon) and for arbitrarily large nuclear system (therefore approximates
the thermodynamical limit) \cite{Chaudhuri2}, it was observed that
results agree with each other under certain conditions. This equivalence
is generally known not to be valid for nuclear systems of finite size.

The main result of this work lies in showing that results from the
canonical and grand canonical models can agree even for finite nuclei.
This equivalence is observed when nuclear multifragmentation leads
to the formation of predominantly nucleons and low mass clusters.
This condition can be achieved either by increasing the temperature
or freeze-out volume of the fragmenting nucleus or source size, or
by decreasing the asymmetry of the source. In fact, when all the four
conditions are satisfied then one can get the best agreement between
the two models. We have confined our study to the observables and
conditions that can be easily accessed by present day experiments.

Specifically we investigate the multiplicity of the fragments leading
to charge and mass distributions from the canonical and grand canonical
distributions under varying conditions and identify the underlying
reasons behind the differences. This led us to identify the conditions
under which results from both the models converge. For example by
comparing charge distributions of fragments obtained from both models
under varying temperature, freeze-out volume, fragmenting source size
and asymmetry, it becomes possible to obtain the conditions under
which the models give rise to similar results.\\
\section{Theoretical formalism}
In this section we describe briefly the canonical and the grand canonical model of nuclear multifragmentation.
The basic output from canonical or grand canonical model is multiplicity
of the fragments. This allows one to obtain the charge or the mass
distribution of the fragments. By multiplicity we mean that the average
number of fragments produced for each proton number $Z$ and neutron
number $N$. Assuming that the system with $A_{0}$ nucleons and $Z_{0}$
protons at temperature $T$, has expanded to a volume higher than
normal nuclear volume and thermodynamical(statistical) equilibrium is reached at this freeze-out condition,
 the partitioning into different composites
can be calculated according to the rules of equilibrium statistical
mechanics.

In a canonical model \cite{Das1}, the partitioning is done such that
all partitions have the correct $A_{0},Z_{0}$ (equivalently $N_{0},Z_{0}$).
The canonical partition function is given by \begin{eqnarray}
Q_{N_{0},Z_{0}} & = & \sum\prod\frac{\omega_{N,Z}^{n_{N,Z}}}{n_{N,Z}!}\end{eqnarray}
where the sum is over all possible channels of break-up (the number
of such channels is enormous) satisfying $N_{0}=\sum N\times n_{N,Z}$
and $Z_{0}=\sum Z\times n_{N,Z}$; $\omega_{N,Z}$ is the partition
function of the composite with $N$ neutrons and $Z$ protons and $n_{NZ}$
is its multiplicity. The partition function $Q_{N_{0},Z_{0}}$ is
calculated using a recursion relation \cite{Das1}. From Eq. (1) and the recursion relation,
the average number of composites is given by \cite{Das1}

\begin{eqnarray}
\langle n_{N,Z}\rangle_{c} & = & \omega_{N,Z}\frac{Q_{N_{0}-N,Z_{0}-Z}}{Q_{N_{0},Z_{0}}}\end{eqnarray}
 It is necessary to specify which nuclei are included in computing
$Q_{N_{0},Z_{0}}$. For $N,Z$ we include a ridge along the line of
stability. The liquid-drop formula gives neutron and proton drip lines
and the results shown here include all nuclei within the boundaries.

In the grand canonical model \cite{Chaudhuri1}, if the neutron chemical
potential is $\mu_{n}$ and the proton chemical potential is $\mu_{p}$,
then statistical equilibrium implies \cite{Reif} that the chemical
potential of a composite with $N$ neutrons and $Z$ protons is $\mu_{n}N+\mu_{p}Z$.
The average number of composites with $N$ neutrons and $Z$ protons
is given by \cite{Chaudhuri1} \begin{eqnarray}
\langle n_{N,Z}\rangle_{gc} & = & e^{\beta\mu_{n}N+\beta\mu_{p}Z}\omega_{N,Z}\end{eqnarray}

The chemical potentials $\mu_{n}$ and $\mu_{p}$ are determined by
solving two equations $N_{0}=\sum Ne^{\beta\mu_{n}N+\beta\mu_{p}Z}\omega_{N,Z}$
and $Z_{0}=\sum Ze^{\beta\mu_{n}N+\beta\mu_{p}Z}\omega_{N,Z}$. This
amounts to solving for an infinite system but we emphasize that this
infinite system can break-up into only certain kinds of species as
are included in the above two equations. We can look upon the sum
on $N$ and $Z$ as a sum over $A$ and a sum over $Z$. In principle
$A$ goes from 1 to $\infty$ and for a given $A$, $Z$ can go from
0 to $A$. Here for a given $A$ we restrict $Z$ by the same drip
lines used for canonical model.

In both the models, the partition function of a composite having $N$
neutrons and $Z$ protons is a product of two parts: one is due to
the the translational motion and the other is the intrinsic partition function of the composite:
\begin{eqnarray}
\omega_{N,Z}=\frac{V}{h^{3}}(2\pi mT)^{3/2}A^{3/2}\times z_{N,Z}(int)
\end{eqnarray}
where $A=N+Z$ is the mass number of the composite and $V$ is
the volume available for translational motion. Note that $V$ will
be less than $V_{f}$, the volume to which the system has expanded
at break up. In general, we take $V_f$ to be equal to three to six times the normal
nuclear volume.  We use $V=V_{f}-V_{0}$ , where $V_{0}$ is the normal
volume of nucleus with $Z_{0}$ protons and $N_{0}$ neutrons. For
nuclei in isolation, the internal partition function is given by $z_{N,Z}(int)=\exp[-\beta F(N,Z)]$
where $F=E-TS$.

For mass number $A=5$ and greater we use the liquid-drop formula
for calculating the binding energy and the contribution for excited
states is taken from the Fermi-gas model. The properties of the composites
used in this work are listed in details in \cite{Mallik1}.\\

\section{Results and Discussions}
\begin{figure}[h]
\includegraphics[width=3.2in,height=2.8in,clip]{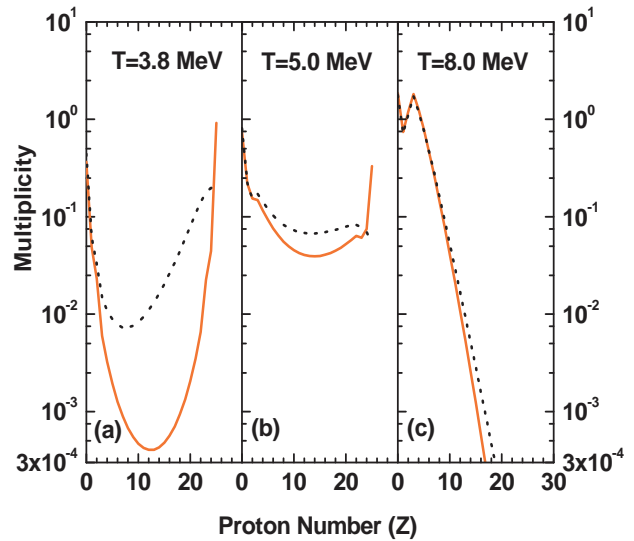}
\caption{(Color online) Total charge distribution of $A_0=60$, $Z_0=25$ system from canonical (red solid lines) and grand canonical model
 (black dotted lines) at same freeze-out volume $V_f=3V_0$ but three different temperatures (a) $3.8$ MeV , (b) $5$ MeV and (c) $8$ MeV.}
\label{fig1}
\end{figure}
\begin{figure}[b]
\includegraphics[width=3.2in,height=2.8in,clip]{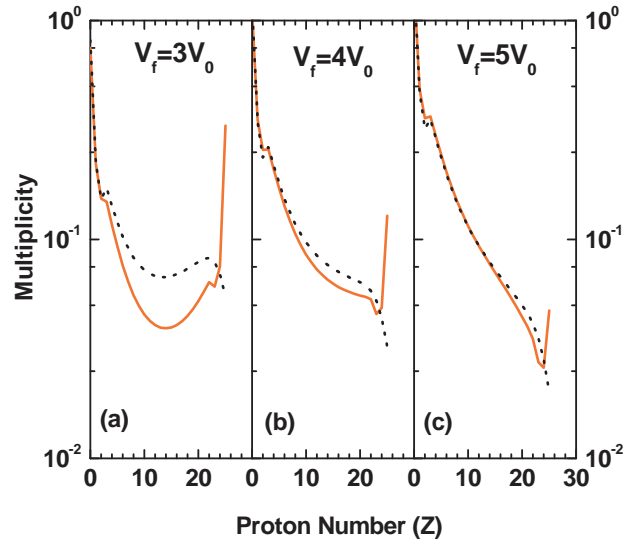}
\caption{(Color online) Total charge distribution of $A_0=60$, $Z_0=25$ system at $T=5.0$ MeV by using canonical (red solid lines) and
grand canonical model (black dotted lines) for three different freeze-out volumes (a) $V_f=3V_0$, (b) $V_f=4V_0$ and (c) $V_f=5V_0$.}
\label{fig2}
\end{figure}
\begin{figure}[t]
\includegraphics[width=3.2in,height=2.8in,clip]{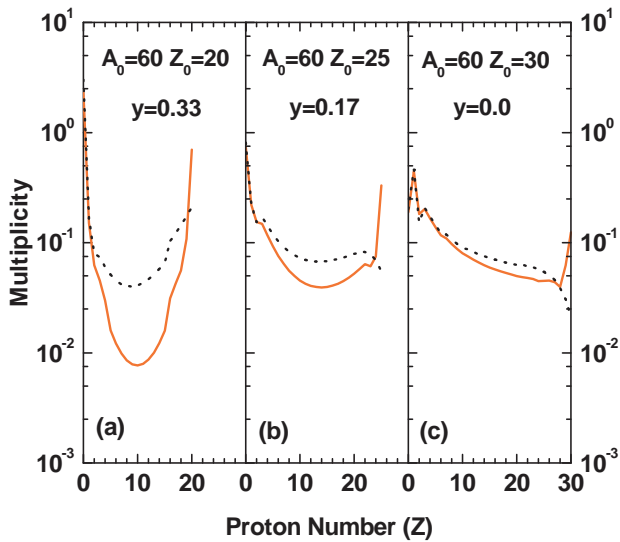}
\caption{(Color online) Total charge distribution at $T=5.0$ MeV and $V_f=3V_0$ from canonical (red solid lines) and grand canonical model
(black dotted lines) of the sources having same $A_0=60$ but different isospin asymmetry (a) $y=0.33$, (b) $0.17$ and (c) $0$.}
\label{fig3}
\end{figure}
\begin{figure}[b]
\includegraphics[width=3.2in,height=2.8in,clip]{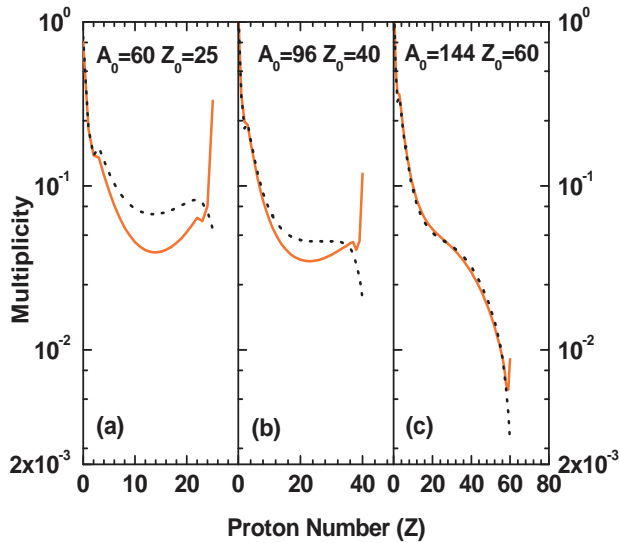}
\caption{(Color online) Total charge distribution at $T=5.0$ MeV and $V_f=3V_0$ by using canonical (red solid lines) and grand canonical
 model (black dotted lines) for three different source sizes $A_0=$ (a)$60$, (b)$96$ and (c)$144$ each having same isospin asymmetry  $y=0.17$.}
\label{fig4}
\end{figure}

We compare the total charge distribution $\langle n_{Z}\rangle=\sum_{N}\langle n_{N,Z}\rangle$
obtained from both the ensembles at different temperatures ($3.8$
MeV, $5$ MeV and $8$ MeV) from disassembly of a particular source
$(Z_{0}=25,A_{0}=60)$ at a fixed freeze-out volume $3V_{0}$ (Fig.
1). The difference in result is maximum at the lowest temperature
$3.8$ MeV where fragmentation is less and the disassembly of the
nucleus results in more of 'liquid-like' fragments or higher mass
fragments. As one increases the temperature, fragmentation increases,
the number of such higher mass fragments decrease (at the expense
of the lower mass ones) and the results from the canonical and grand
canonical ensembles begin to converge. This is easily seen at the
two higher temperatures. At $8$ MeV the results from both the ensembles
are very close to each other since fragmentation is maximum at this
temperature, the nucleons and the lower mass fragments dominating
the distribution.

The effect of increasing the freeze-out volume (decreasing the density)
is equivalent to that of increasing the temperature and this is seen
in Fig. 2. Here we have repeated the same calculation for the same
source at $T=5$ MeV for three different freeze-out volumes. It is
seen that results from both the ensembles agree with each other as
one increases the freeze-out volume when the nucleus fragments more
into smaller pieces. Similar effect is also seen if we vary the source
asymmetry $y=(N_{0}-Z_{0})/(N_{0}+Z_{0})$ keeping the temperature
fixed  $5$ MeV, freeze-out volume at $3V_{0}$ and source size at $A_0=60$. Fig 3
shows the charge distribution for three nuclei having $y$ = 0.33,
0.17 and 0 respectively. We observe that the difference in results
between both the ensembles is maximum when the asymmetry is more (Fig
3(a)) and the difference is least for the symmetric nucleus (Fig 3(c)).

The reason behind the differences is the same as that in the case
of temperature variation. When the nucleus is more asymmetric, fragmentation
(breaking of the nucleus) is less and the fraction of higher mass
fragments is more as compared to the more symmetric case which will
be shown later. This effect is also seen if we keep both temperature,
freeze-out volume and the asymmetry parameter fixed but increase the
source size(mass) as shown in Fig. 4. The difference in result between both
the ensembles is maximum when the source size is minimum as expected
and the results become close to each other for a large nucleus. We
can say that the nucleus fragments more and more as one increases
the source size (keeping other parameters fixed) and the effect is
similar to that of increasing the temperature keeping the source size
fixed. \\

In order to investigate the effect more, we have calculated the ratio
(normalized) of higher mass fragments formed to that of the total
number of fragments (total multiplicity). The fragment whose size
is more than 0.8 times $A_{0}$ (more than $80\%$ of the source in
size) are considered as higher mass fragments i.e. the ratio is defined
as \begin{eqnarray}
\eta=\frac{\sum_{A>0.8A_{0}}^{A_{0}}\langle n_{N,Z}\rangle}{\sum_{A=1}^{A_{0}}\langle n_{N,Z}\rangle}\end{eqnarray}
 This criteria of choosing the higher mass fragments is not very rigid
and can be relaxed. We have checked that even if we make it 0.75 or
0.85 instead of 0.8 the trend of the results remain same. We have done this calculation in both canonical and grand canonical models and the results are similar. We have shown the results in Fig 5 from the grand canonical model. In Fig 5.a
we show the variation of this ratio as a function of temperature (keeping
source size, freeze-out volume and asymmetry fixed) and it is seen
that the ratio decreases with  increase in $T$. This shows that for a source with
lower values of $T$, the fraction of higher mass fragments formed
as a result of fragmentation is more as compared to those with higher
$T$ values. We emphasize that the difference in the charge distributions
from the canonical and grand canonical ensembles is mainly caused
by the presence of the higher mass fragments in the distribution.
The lesser is the fraction of the higher mass fragments, the deviation
in results between both the ensembles will be less and this is exactly
what we saw in Fig. 1. Similar effect is seen when one plots this
ratio (Fig 5.b) as a function of $V_{f}/V_{0}$ keeping other parameters
fixed. It is clearly seen that with increase in the freeze-out volume,
the fraction of higher mass fragments decrease and this causes the
results between both the ensembles to be very close when $V_{f}$
is maximum as shown in Fig. 2.
\begin{figure}[b]
\includegraphics[width=3.2in,height=3.2in,clip]{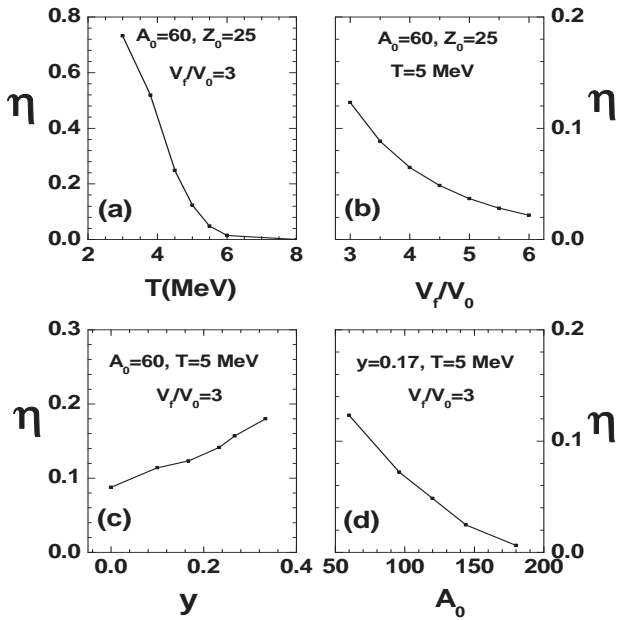}
\caption{Variation of $\eta$ with (a) temperature, (b) freeze-out volume, (c) isospin asymmetry and (d) source size from grand canonical model.}
\label{fig5}
\end{figure}
We also plot $\eta$ as a function of the asymmetry parameter
$y$ of the source, the source size ($A_{0}=60$), temperature (5
MeV) and freeze-out volume ($3V_{0}$) being kept fixed and it is
seen that the ratio increases with $y$ (Fig 5(c)). So here we observe
that the less is the asymmetry of the source, less is the number of large
fragments and hence fragmentation of the nucleus is more. In this
scenario, when the nucleus is more symmetric the results from the
two ensembles agree to a very good extent than when the nucleus is
less symmetric as seen in Fig 3. The same effect is seen (Fig 5(d))
if one increases the source size keeping the other parameters fixed
and we assert that the effect of increasing the source size is similar
to that of increasing the temperature or freeze-out volume or decreasing
the asymmetry of the source as far as convergence between both the
ensembles is considered. What we wish to convey is that the differences
in results between the canonical and the grand canonical ensemble
is mainly because of the presence of the higher mass fragments in
the fragmentation of a nucleus. If the conditions are such that the
fragmentation is more and there are only lower mass clusters, then
the results from both the ensembles agree to a much better extent. The same condition is also valid for convergence between microcanonical and canonical ensembles where energy plays the role of the extensive variable instead of the total number of particles. The more the nucleus disintegrates, the less will be the fluctuation in energy and better will be the convergence between the microcanonical and the canonical ensembles.\\

\section{Summary and Conclusion}
This Letter analyzes the charge distributions
of fragments formed in nuclear multifragmentation in both canonical
and grand canonical versions of the multifragmentation model. Both models
are typically used to study experimental data from heavy-ion collisions
at intermediate energies. We have shown that results from both models
are in agreement for finite nuclei provided the nucleus fragments
predominantly into nucleons and low mass clusters. We have seen that
this condition is achieved under certain conditions of temperature,
freeze-out volume, source size and source asymmetry. The main message
that we wish to convey in this work is that while canonical and grand
canonical models have very different underlying physical assumption,
the results from both models can be in agreement with each other provided
the contribution of higher mass fragments in nuclear disassembly is
insignificant. This condition can be achieved either by increasing
the the temperature or freeze-out volume of the fragmenting nucleus
or by increasing the source size, or by decreasing the asymmetry of
the source. In fact when all these four conditions are satisfied then
one obtains the best convergence between the two models. On the other
hand, when the temperature and freeze-out volume are low, nucleus
is small and more asymmetric then fragmentation of the nucleus is
least; in these cases higher mass fragments dominate the distribution
and the results from both the ensembles will be very different We would like to add that the convergence between the microcanonical and the canonical ensemble will also be achieved under the similar conditions as those between the canonical and the grand canonical ensembles.

\section{Acknowledgement}
We would like to thank Prof. S. Das Gupta (Mcgill University) for introducing us to this subject.

\end{document}